\documentclass[10pt
,aps
,pra
,amsmath
,amssymb
,amsfonts
,twocolumn
,notitlepage
,longbibliography
]{revtex4-2}

\usepackage{graphicx}% Include figure files
\usepackage{dcolumn}% Align table columns on decimal point
\usepackage{bm}     % bold math
\usepackage{braket}
\usepackage[hidelinks]{hyperref}% add hypertext capabilities
\usepackage{scalerel}
\allowdisplaybreaks
\usepackage{lipsum}  
\usepackage{mathrsfs}
\usepackage{bbold}
\usepackage{upgreek}
\usepackage{dsfont}
\usepackage{amsmath}
\usepackage{booktabs}
\usepackage{siunitx}
\newcommand\numberthis{\addtocounter{equation}{1}\tag{\theequation}}
\begin{document}

%\preprint{APS/123-QED}

\title{High-throughput assessment of defect-nuclear spin register controllability for quantum memory applications}% Force line breaks with \\

\author{Filippos Dakis}
 \email{dakisfilippos@vt.edu}
\affiliation{Department of Physics, Virginia Tech, Blacksburg, Virginia 24061, USA}
\affiliation{Virginia Tech Center for Quantum Information Science and Engineering, Blacksburg, Virginia 24061, USA}%
\author{Evangelia Takou}%
 \email{etakou@vt.edu}
\affiliation{Department of Physics, Virginia Tech, Blacksburg, Virginia 24061, USA}
\affiliation{Virginia Tech Center for Quantum Information Science and Engineering, Blacksburg, Virginia 24061, USA}%
\author{Edwin Barnes}%
 \email{efbarnes@vt.edu}
\affiliation{Department of Physics, Virginia Tech, Blacksburg, Virginia 24061, USA}
\affiliation{Virginia Tech Center for Quantum Information Science and Engineering, Blacksburg, Virginia 24061, USA}%
\author{Sophia E. Economou}%
 \email{economou@vt.edu}
\affiliation{Department of Physics, Virginia Tech, Blacksburg, Virginia 24061, USA}
\affiliation{Virginia Tech Center for Quantum Information Science and Engineering, Blacksburg, Virginia 24061, USA}%
\date{\today}% It is always \today, today,

\begin{abstract}
Quantum memories play a key role in facilitating tasks within quantum networks and quantum information processing, including secure communications, advanced quantum sensing, and distributed quantum computing. Progress in characterizing large nuclear spin registers coupled to defect electronic spins has been significant, but selecting memory qubits remains challenging due to the multitude of possible assignments. Numerical simulations for evaluating entangling gate fidelities encounter obstacles, restricting research to small registers, while experimental investigations are time-consuming and often limited to well-understood samples. Here we present an efficient methodology for systematically assessing the controllability of defect systems coupled to nuclear spin registers. We showcase the approach by investigating the generation of entanglement links between silicon monovacancy or divacancy centers in SiC and randomly selected sets of nuclear spins within the two-species ($^{13}$C and $^{29}$Si) nuclear register. We quantify the performance of entangling gate operations and present the achievable gate fidelities, considering both the size of the register and the presence of unwanted nuclear spins. We find that some control sequences perform better than others depending on the number of target versus bath nuclei. This efficient approach is a  guide for both experimental investigation and engineering, facilitating the high-throughput exploration of suitable defect systems for quantum memories.
\end{abstract}

%\keywords{Suggested keywords}%Use showk eys class option if keyword
\maketitle

%\tableofcontents

\section{\label{Sec:Introduction} Introduction}
Solid-state defect systems are a leading platform for quantum networks, quantum sensing, and other quantum information processing tasks. Defects in diamond such as nitrogen-vacancy (NV) centers \cite{Pompili2021, Bernien2013, Kalb2017}, SiV centers \cite{Nguyen2019_1, Nguyen2019_2}, and SnV centers~\cite{Iwasaki2017} have been intensely studied and used for milestone demonstrations of building blocks for quantum networks. Similarly, various defects in silicon carbide (SiC) are explored for such tasks due to the host material’s attractive properties and the more desirable emission frequencies compared to diamond defects. Among SiC defects, silicon vacancies \cite{Babin2021,Nagy2019,Widmann2014} and neutral divacancy centers \cite{Bourassa2020} are especially promising.\\ 
\indent Considering the vast space of possible defects in a wide range of host materials, the community is taking steps toward a more systematic search for new defects with desirable properties using computational tools such as DFT~\cite{Alkauskas2019,Broberg2023}. The latter predict optoelectronic and kinematic properties, which, while important, do not suffice in determining the suitability of a given defect for a particular quantum information processing task.\\
\indent The most fundamental missing element in such defect surveys is the evaluation of the \textit{operational performance} of the defect as a qubit in a larger quantum register, typically composed of spinful isotopes in the host material. For example, in network applications, the electronic spin functions as the communication qubit due to its spin-photon interface, while nearby nuclear spins can serve as long-lived quantum memories needed for quantum repeaters \cite{Briegel1998, Brand2020,Bergeron2020}. In the case of diamond for instance, the low but nonzero concentration of $^{13}$C nuclei, together with their long coherence times, allows for the development of quantum memories for information storage and buffering. Such memory qubits are also often necessary for entanglement distillation \cite{Kalb2017} and quantum error correction protocols \cite{Waldherr2014}, while they can also serve as a useful resource for quantum sensing \cite{Abobeih2019,vandeStolpe2024, Giovannetti2004, Zaiser2016}.\\
\indent However, assessing and controlling nuclear spin memories in solid-state spin defect platforms faces challenges on two fronts: (i) The nuclear spins are randomly situated at distant lattice sites, resulting in random hyperfine (HF) interactions that are weak compared to the defect spin's dephasing rate. (ii) The HF interactions between nuclear spins and the electronic spin defect are always present (they cannot be switched off). Thankfully, dynamical decoupling (DD) pulse sequences offer a solution to both challenges \cite{Taminiau_2012}. In principle, all nuclear spins except a selected target one can be decoupled from the defect by tuning parameters associated with these DD sequences, particularly the inter-pulse spacings and the number of iterations of a basic sequence unit. By varying these control knobs, different nuclear spins across the total register can be selected. To date, impressive experiments have been conducted to characterize the register and demonstrate entangling quantum gates between defect and nuclear spin qubits \cite{Taminiau_2012,Taminiau2014,Abobeih2019,Bradley2019,Nguyen2019_1,Nguyen2019_2,Bourassa2020}. Moreover, this approach has initiated first steps towards distributing entanglement across a network of a few quantum nodes \cite{Pompili2021, Humphreys2018}, implementing quantum repeater protocols \cite{Bradley2019}, performing entanglement distillation \cite{Kalb2017}, or realizing error-correction schemes \cite{Taminiau2014, Cramer2016, Abobeih2022}. Very recently, entanglement between nuclear spin qubits coupled to two SiV centers in diamond at opposite ends of a 40 km fiber was demonstrated \cite{Knaut2024}.\\
\indent Despite the significant progress in characterizing large nuclear spin registers \cite{vandeStolpe2024}, selecting memory qubits from these registers remains inherently difficult due to the combinatorially large number of possible assignments. An important metric in this selection process is the quality of the gates for a given assignment. Unfortunately, numerical simulations to evaluate gate fidelity face significant hurdles, as they require solving for the time-evolution of the many-body system, rendering such calculations prohibitively difficult for large registers. Consequently, only small-sized registers could be simulated so far. Although experimental investigations could partially alleviate this challenge, they are time-consuming, and academic labs often focus their investigations on a few well-understood and characterized samples and are not yet as concerned about questions of yield and scale.\\
\indent In this paper, we address this challenge by establishing an efficient method for assessing the controllability of the total system consisting of the defect spin coupled to an arbitrary multi-nuclear spin register. We do this by leveraging a recent formalism for characterizing multi-spin entanglement \cite{Takou2023} that completely sidesteps the need to simulate many-body dynamics—a notoriously resource-intensive and time-consuming computational task. Our method enables one to rapidly and systematically identify suitable nuclear spins to use as memories, as well as optimal control sequences for manipulating them with high fidelity. While our method is general and can be applied to any defect, we showcase our approach with silicon monovacancy and divacancy 
defects in silicon carbide coupled to two species of nuclear spins, \textsuperscript{13}C and \textsuperscript{29}Si. These defects are especially promising for networks due to the fact that they emit at or near telecom bands. We determine how electron-multi-nuclear gate fidelities depend on the number of target versus bath nuclei. We compare the performance of several leading DD sequences in this task, finding that the relative performance depends on the sizes of the nuclear spin memory register and the nuclear spin bath. We further show that the sign difference in the gyromagnetic ratios of the two nuclear spin species is beneficial for achieving high-fidelity multi-spin operations. Our results constitute an important step in comparing the utility of different defect systems by allowing memory controllability to be included systematically in such comparisons.\\ 
\indent The paper is organized as follows. In Sec.~\ref{Sec:Background}, we review the case of a defect spin coupled to a single nuclear spin and driven by $\pi$-pulse sequences. In Sec.~\ref{Sec:Multiple_nuclei}, we generalize this to the case of multiple nuclear spins. We first discuss the metric we use to quantify entanglement, and we then apply this to selectively entangle a subset of nuclei in the case of a defect in SiC coupled to $^{13}$C and $^{29}$Si nuclei. In Sec.~\ref{sec:Results}, we study the dependence of the gate fidelity on the size of the multipartite entangled state, the size of the bath, and the type of control sequence used. We summarize our conclusions in Sec.~\ref{Sec:Conclusion}. 

\section{\label{Sec:Background} Single Nuclear Spin Coupled To Defect and dynamical decoupling}
\begin{figure}
    \includegraphics[width=1\linewidth,keepaspectratio]{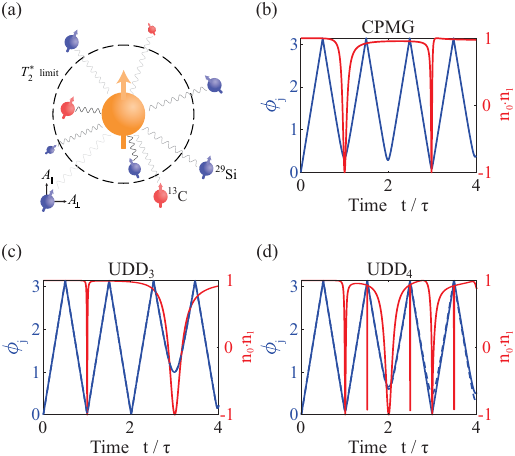}
    \caption{\label{fig:Schematic_and_DotProducts} (a) Schematic depicting a single silicon vacancy coupled to nuclear spins through HF interaction. The spheres denote both nuclear (\textsuperscript{13}C, \textsuperscript{29}Si) and electronic spins. Additionally, we illustrate the transition distance between the strongly coupled and weakly coupled regimes, depicted as the $T_2^*$ limit (black dashed line). In (b)-(d) we plot the rotation angles $\phi_j$ (solid and dashed blue lines) and the dot product $\mathbf{n}_0\cdot \mathbf{n}_1$ (red color) of nuclear axes as a function of the unit time $\tau_k$ of CPMG, UDD\textsubscript{3} and UDD\textsubscript{4}, respectively. For CPMG and UDD\textsubscript{3}, it holds that $\phi_0=\phi_1$ and thus the blue lines (solid and dashed) are on top of each other, while in UDD\textsubscript{4} the two angles are distinguishable. For the nuclear spin, we set $(\omega_L,\, A_{\|},\,A_{\bot}) = 2\pi\cdot(100,\,100,\,80)\, \rm kHz$, and the electron’s spin projections are $(s_0,\,s_1) = (1/2,\, 3/2)$. The time is normalized to the first order resonance times $(\tau_{\rm CPMG},\tau_{\rm UDD_3},\tau_{\rm UDD_4}) = (4.6487,\,4.6444,\, 4.6495)\, \rm \mu s $.}
\end{figure}

Reference \cite{Taminiau_2012} demonstrated that by selecting the pulse intervals of a DD sequence to satisfy a specific resonance condition dictated by the hyperfine (HF) couplings, it is possible to rotate a desired nuclear spin conditionally on the electronic spin state. This is possible because DD sequences can dynamically modify the effective electron-nuclear HF interaction, enabling the coupling of a specific nuclear spin to the electron spin while decoupling others. Well-known examples of DD sequences that are widely used in the quantum information science community include the Carr-Purcell-Meiboom-Gill (CPMG) \cite{Carr1954, Meiboom1958, deLange2010} and Uhrig (UDD) \cite{Uhrig2007} sequences. In this section, we review existing results for single nuclear spin control via electronic spin driving.\\
\indent As a first step, we review how one can create electron-nuclear spin entanglement using repeated DD pulses sequences. For instance, one can use the CPMG sequence on the electron and repeat the basic unit $N$ times, i.e., $(\tau/4-\pi-\tau/2-\pi-\tau/4)^N$, where $\tau$ is the duration of the unit, and $\pi$ represents the application of an instantaneous $\pi$-pulse. These $\pi$-pulses are implemented experimentally via a microwave (MW) drive to induce transitions between electronic spin states directly. In practice, the $\pi$-pulses have a finite duration and amplitude, and they can be generated using a vector source \cite{Bradley2021} where their features such as frequency, duration, and amplitude, are preset by an arbitrary waveform generator, while their pulse shapes can be Hermite envelopes \cite{Bradley2019, Vandersypen2005}.
\\\indent The Hamiltonian for a single electronic spin in SiC interacting with a single nuclear spin $(I = 1/2)$ is given by~\cite{Freeman1967-ut} 
\begin{equation}
\begin{split}
     H &= \frac{\omega_L}{2}\mathbb{1}\otimes\sigma_{z} + \frac{A_{\|}}{2}Z_e\otimes\sigma_{z} + \frac{A_{\perp}}{2}Z_{e}\otimes\sigma_{x}\\
     &= \sigma_{00}\otimes H_{0} + \sigma_{11}\otimes H_{1}\, ,
\end{split}
\label{eq:Hamiltonian_1}
\end{equation}
where $\omega_{L}$ is the Larmor frequency of the nuclear spin, $\sigma_{i}$, with $ i\in\{x,y,z\}$, are the Pauli matrices, and $A_{\|}$ $(A_{\perp})$ is the parallel (perpendicular) component of the HF interaction. The electronic spin operator $Z_e$ is defined as $Z_e = s_0\sigma_{00} + s_1\sigma_{11}$, where $\sigma_{jj}=\ket{j}\bra{j}$ are the electronic spin projection operators, with $\ket{j=0}$ and $\ket{j=1}$ being the two levels of the electronic spin multiplet used to define the two-dimensional qubit space, while $s_{j}$ is the corresponding spin projection quantum number. For instance, NV centers in diamond have total spin $S=1$ and the logical qubit states are $\ket{s_0=0}$ and $\ket{s_1=-1}$ \cite{Bernien2013}, while the negatively charged silicon vacancy in SiC has $S=3/2$ and the qubit states are typically defined to be $\ket{s_0 = 1/2}$ and $\ket{s_1 = 3/2}$ \cite{Nagy2019}, or $\ket{s_0 = -1/2}$ and $\ket{s_1 = -3/2}$ \cite{Widmann2014}. Moreover, $H_{j} = 1/2\left[(\omega_{L} + s_{j}A_{\|})\sigma_{z} + s_{j}A_{\perp}\sigma_{x}\right]$ is the Hamiltonian experienced by the nuclei, which is conditional on the state of the electronic spin. Using the above Hamiltonian, one can show that the evolution operator, after one unit of the pulse sequence, can be written as \cite{Taminiau_2012, Takou2023}
\begin{equation}
\label{eq:Propagator_single_pulse}
\begin{split}
    U = \sigma_{00}\otimes R_{\mathbf{n}_{0}}(\phi_{0}) + \sigma_{11}\otimes R_{\mathbf{n}_1}(\phi_{1})\, ,
\end{split}
\end{equation}
where $R_{\mathbf{n}_j}(\phi_{j}) = e^{-i\phi_j/2\bm{\sigma}\cdot\mathbf{n}_j}$ are the two conditional nuclear spin evolution operators defined by two rotation axes $\mathbf{n}_{j}$ and two angles $\phi_{j}$. Both $\mathbf{n}_j$ and $\phi_j$ depend on the electronic spin state and the chosen pulse sequence. The potential of creating controlled gates with the electron as the control qubit and the nuclei as the target qubits is already evident in Eq.~\eqref{eq:Propagator_single_pulse}. For instance, in the case where $\phi_1=\phi_0 = \pi/2$ and $\mathbf{n}_0 = -\mathbf{n}_1 = \mathbf{x}$, the evolution operator yields $ U = \mathrm{CR}_{\mathrm{x}}(\pi/2) = \sigma_{00}\otimes R_{\mathbf{x}}(\pi/2) + \sigma_{11}\otimes R_{\mathbf{-x}}(\pi/2)$, which is equivalent to {\footnotesize $\mathrm{CNOT}$} up to local rotations.\\
\indent To leverage this conditional Hamiltonian to create electron-nuclear entanglement, the electron spin must be prepared in a superposition state, e.g., $\ket{+} = (\ket{0} + \ket{1})/\sqrt{2}$, while the nuclear spin can be initialized in the state $\ket{0}$. In certain cases, the subsequent application of $N$ DD sequence units makes the two nuclear rotation operators, $R_{\mathbf{n}_j}(N\phi_j)$, differ. This can happen for example, if the unit time, $\tau$, is chosen such that the nuclear rotation axes are antiparallel, i.e., $\mathbf{n}_0\cdot\mathbf{n}_1 = -1$. At this time, the probability, $P_+$, that the initial state $\ket{+}$ of the electron is preserved reaches the minimum value, signifying the creation of an electron-nuclear spin Bell-pair. This probability (or coherence function) is calculated as $P_+ = (1+M)/2$, with $M = \frac{1}{2}\hbox{Re tr}[R_{\mathbf{n}_0}(N\phi_0)R_{\mathbf{n}_1}^{\dagger}(N\phi_1)]$\cite{Taminiau_2012, Takou2023}. For instance, in the case of CPMG pulses, it holds that $\phi_0 = \phi_1 = \phi$ and $M = 1 - \sin^{2}(N\phi/2)(1-\mathbf{n}_0\cdot{\mathbf{n}}_1)$, where $N$ is the number of the applied unit sequences. This last expression reveals that for antiparallel axes and accumulated phase $N\phi = \pi/2$ the resulting conditional rotation leads to a dip in the electron's coherence function down to the value $P_+ = 1/2$. The resonance times at which this entanglement occurs can be calculated using the explicit formulas for the nuclear operators $R_{\mathbf{n}_j}(\phi_j)$, setting $\mathbf{n}_0\cdot{\mathbf{n}}_1 = -1$, and solving ${\rm tr}(R_{\mathbf{n}_0}R_{\mathbf{n}_1}^\dagger)/2 = \cos(\phi_0)\cos(\phi_1) + \mathbf{n}_0\cdot\mathbf{n}_1\sin(\phi_0)\sin(\phi_1)$ for time $\tau$, where $R_{\mathbf{n}_j}$ and $\phi_j$ depend on the pulse sequence and the unit time $\tau$. One can show that for CPMG, UDD\textsubscript{3} and UDD\textsubscript{4} sequences, the resonance times are given by \cite{Taminiau_2012, Takou2023}
\begin{equation}
\label{eq:resonance_times}
    \tau_{k} \simeq \frac{4\pi(2k-1)}{\omega_0 + \omega_1}\,,
\end{equation}
where $\omega_j = \sqrt{(\omega_L + s_jA_{\|})^2 + (s_jA_{\perp})^2}$, and $k \in \mathds{Z}^+$ is the resonance order. As reported in \cite{Dong2020}, the UDD\textsubscript{4} sequence has additional resonance times located at $\tau_k \simeq 8\pi(2k-1)/\tilde{\omega}$, see Fig,~{\ref{fig:Schematic_and_DotProducts}(d)}. These expressions for $\tau_k$ are approximate and valid for $\omega_L\gg A_{\|},A_{\perp}$.\\
\indent In general, when the resonance condition is satisfied ($\mathbf{n}_0\cdot\mathbf{n}_1 = -1$), a single DD unit creates a small angle of rotation, as shown in Figs.~\ref{fig:Schematic_and_DotProducts}(b)-(d), and thus a small amount of entanglement. A larger amount of electron-nuclear spin entanglement can be achieved by iterating the unit sequence $N$ times to accumulate the desired angle of rotation $N\phi_j$. In Fig.~\ref{fig:Schematic_and_DotProducts}, we present the angle of rotation versus the unit time for the CPMG, UDD\textsubscript{3}, and UDD\textsubscript{4} sequences. Sequences with an odd number of pulses in the basic unit need to be repeated twice to ensure the electron experiences an even number of $\pi-$pulses and returns to its initial state \cite{Takou2023}. In the case of CPMG and UDD\textsubscript{3} the rotation angles are the same, i.e., $\phi_0=\phi_1$. A straightforward way to create a large amount of entanglement is to set the unit time equal to a resonance time ($\tau=\tau_k$), so that the dot product is always $\mathbf{n}_0\cdot\mathbf{n}_1 = -1$,  and apply $N$ sequences such that the accumulated angle is $N\phi\simeq \pi/2$. This combination leads to a perfect entangling gate $\mathrm{CR}_{\mathbf{x}}(\pi/2)$, which is equivalent to {\footnotesize{CNOT}} up to single qubit rotations.\\
\indent Although the time intervals between the $\pi$-pulses are given by the same formula for all UUD\textsubscript{n} sequences (see Appendix~\ref{App_1}), the rotation angles are not equal for some of them. The UDD\textsubscript{4} sequence leads to a more complicated evolution of the nuclear spin in which the rotation angle depends differently on the state of the electron. This in turn leads to a nontrivial dependence of the dot product of its rotation axes on $N$. Thus, even if one fixes a resonance time for the basic UDD\textsubscript{4} unit, the nuclear rotation axes can switch from antiparallel to parallel for values of $N$ where $\phi_0$ and $\phi_1$ become equal \cite{Takou2023}. In this case, with parallel axes $(\mathbf{n}_0\cdot\mathbf{n}_1 = 1)$ and equal angles of rotation, the nuclear spin undergoes an unconditional rotation and, thus, no entanglement can be created.\\
\indent As we show later on,  perfect entanglement can also be achieved by easing the restriction of the unit time being equal to a resonance time, as long as the dot product remains negative, $\mathbf{n}_0\cdot\mathbf{n}_1<0$. This is indeed very important since it makes it possible to entangle the electron with several nuclear spins (with different HF interactions) simply by tuning the pulse spacing $\tau$ to such a ``multi-nuclear resonance" time. Furthermore, the above feature can be combined with the fact that $\pi$-pulse sequences can also average out the interactions of the electron with unwanted spins, ensuring some degree of selectivity with a target set of spins.

\section{\label{Sec:Multiple_nuclei} Entanglement in Defect-Multi-nuclear spin systems}
We now move on to consider a defect electronic spin coupled to a multi-nuclear spin register. We first discuss metrics for gauging the ability of HF interactions and DD sequences to create entanglement in these systems. We then discuss how to apply these metrics to select which nuclear spins to entangle with the electronic spin and which to decouple. We focus on the case of defects in SiC coupled to two nuclear species ($^{13}$C and $^{29}$Si) for concreteness.

\subsection{Quantifying entanglement in electron-multi-nuclear spin systems\label{sec:Quantifying_Entanglement}}
\begin{figure}
    \includegraphics[width=1\linewidth,keepaspectratio]{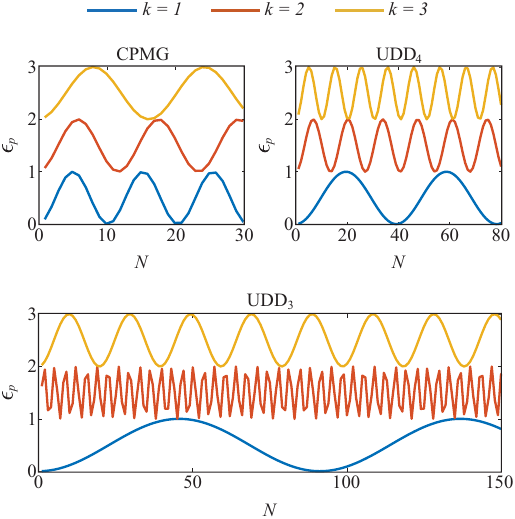}
    \caption{\label{fig:Resonance_Order} Scaled entangling power as a function of resonance order ($k=1$ blue lines; $k=2$ red lines; $k=3$ yellow lines) and the number of repetitions $N$ of the CPMG (left), UDD\textsubscript{4} (right), and UDD\textsubscript{3} (bottom) units. For the nuclear spin, we set $(\omega_L,\, A_{\|},\,A_{\bot}) = 2\pi\cdot(-300,\,60,\,30)\, \rm kHz$, and the electron’s spin projections are $(s_0,\,s_1) = (1/2,\, 3/2)$. The resonance times are optimized around the analytical resonance time, for CPMG $(\tau_1,\,\tau_2,\,\tau_3) = (4.1378,\, 12.354,\,20.6201)\,\rm \mu s$, for UDD\textsubscript{3} $(\tau_1,\,\tau_2,\,\tau_3) = (4.1360,\, 12.404,\,20.6170)\,\rm \mu s$, and for UDD\textsubscript{4} $(\tau_1,\,\tau_2,\,\tau_3) = (4.1393,\, 12.3743,\,20.26451)\,\rm \mu s$. The scaled entangling power takes values $\epsilon_p \in [0,1]$, and the offsets in the graphs are used for illustration purposes only.}
\end{figure}

Before we develop protocols for controlling and entangling multiple nuclear spins, we first have to discuss how we quantify entanglement in multi-nuclear spin registers. We begin with considering the entangling power for a defect coupled to a single nuclear spin and then extend the formalism to multiple nuclear spins by employing the concept of one-tangles \cite{Linowski2020}. As will become clear in Sec.~\ref{sec:Results}, these quantities will guide our design towards selective multi-nuclear spin entangling gates. \\
\indent The entangling power is a general property of logical gate operations that disregards the details of a gate and focuses solely on its entanglement capabilities \cite{Zanardi_2000}. The entangling power of a two-qubit operator can be expressed as \cite{Balakrishnan_ep}
\begin{equation}
\label{eq:entangling_power}
    \epsilon_p = 1-|G_1|\,,
\end{equation}
where the function $G_1$ is a Makhlin invariant (see Appendix~\ref{App:Makhlin_invariants}) whose explicit form in the present context of electron-nuclear entanglement is
\begin{equation}
\label{eq:G1}
\begin{split}
    G_1 = \bigg(\cos\frac{\phi_0(N)}{2}&\cos\frac{\phi_1(N)}{2}\\
    &+ n_{01}\sin\frac{\phi_0(N)}{2}\sin\frac{\phi_1(N)}{2}\bigg)^2\,.
    \end{split}
\end{equation}
Here, we use the fact that for any given $\pi$-pulse sequence repeated $N$ times, the electron-nuclear evolution operator maintains the structure described in Eq.~\eqref{eq:Propagator_single_pulse}, with the only difference being that $\phi_{j}$ is now replaced by the total rotation angle $\phi_{j}(N)$. In Eq.~\eqref{eq:entangling_power}, we discarded an overall factor of $2/9$ for convenience, and thus the entangling power takes values in the range $0\leq\epsilon_p\leq1$. 
It is evident that when $G_1$ becomes 0, the entangling power is maximized, reaching its peak at $\epsilon_p = 1$ for the two-qubit case. The minima of $G_1$ occur at $N =(2k+1)\pi/(\phi_0 + \phi_1)$; however, $N$ is an integer and so the value of this expression must be rounded in general. Notice that $G_1$ can also be zero for other $N$ values, provided that $\mathbf{n}_0 \cdot \mathbf{n}_1 \leq0$. In Fig.~\ref{fig:Resonance_Order} we plot the scaled entangling power, $\epsilon_p$, for a single nuclear spin coupled to the electron versus the number of iterations and the resonance order $k$. We see that the frequency of the maxima of $\epsilon_p$ depends on the pulse sequence and the resonance order, an expected behavior since these features directly affect the rotation angle per iteration. \\
\indent To quantify the distribution of entanglement within the entire system, consisting of one electronic and many nuclear spins, we expand the notion of entangling power beyond two qubits. To that end, we utilize the one-tangles \cite{Coffman2000} which measure the overall entanglement within a state by assessing all possible bipartitions of the system. By virtually partitioning the total system into subsystems, we quantify the level of correlations between these subsystems (this is also known as bipartition entanglement). We choose this metric to quantify the entanglement of our gate, and for each bipartition we separate only one qubit (either electron or nuclear spin) from the rest of the system. Note that one-tangles exclusively convey information about the system's entanglement capacity, and they cannot differentiate between states belonging to different categories (e.g., W states versus GHZ states in the $n$-partite case, $n\geq 3$) \cite{Linowski2020}. We find one-tangles to be a convenient metric in our system since we are interested in its entanglement capacity, rather than generating specific entangled states.\\
\indent For the sake of completeness, we note that for a pure state $\ket{\psi}$, the one-tangle is defined by the generalized concurrence \cite{Linowski2020}
\begin{equation}\
    \label{eq:one-tangle_definition}
    \uptau_{g|g'}(\ket{\psi}) := 2(1 - {\rm tr}[\rho_{g'}^2])\,,\quad \rho_{g'}={\rm tr}_g[\ket{\psi}\bra{\psi}]\, ,
\end{equation}
where $g|g'$ denotes the bipartition or the splitting of the Hilbert space. Equation~\eqref{eq:one-tangle_definition} bounds one-tangles in the range $[0, 2-2/{\rm min}(d,d')]$, where $d$ ($d'$) is the dimension of partition $g$ ($g'$).\\
\indent The present form of Eq.~\eqref{eq:one-tangle_definition}) is not very effective for measuring the entanglement of multi-nuclear operations because it relies on the initial state. To address this, we need to consider an average across all initial states. Specifically, we employ the bipartition entangling power, which is the average of the one-tangle across all initial product states. According to Ref.~\cite{Linowski2020}, this is calculated by averaging over single-qubit unitaries $U_i$ applied to initial product state $\ket{\psi_0}$, denoted as $\ket{\Psi}:=\ket{\psi_0}^{\otimes_i} = U^{\otimes_i}_i\ket{\psi_0}^{\otimes_i}$, resulting in $\epsilon_{g|g'}(U):=\braket{\uptau_{g|g'}(U\ket{\Psi})}_{U_i}$. Here, the index $i$ ranges from 1 to $n$, where $n$ represents the total number of qubit systems. Ordinarily, this ``one-tangling power" is difficult to calculate explicitly for a generic many-body Hamiltonian. However, as first shown in Ref.~\cite{Takou2023}, it can be analytically computed for the type of electron-nuclear central spin system that we have in defect systems. Although $\epsilon_{g|g'}(U)$ characterizes the bipartite entanglement generated by the unitary $U$ and is not a state-dependent property, for brevity we will still refer to it as a ``one-tangle" in what follows.\\
\indent Before discussing the explicit expression for the one-tangle, we first introduce the Hamiltonian of a single electronic spin coupled to multiple nuclear spins. 
%In the case of two nuclear spins, and neglecting direct inter-nuclear interactions, the Hamiltonian reads
%\begin{equation}
%    \label{eq:Hamiltonian_two_nuclear_spins}
%\begin{split}
%    H =& \frac{\omega_L^{(1)}}{2}\mathbb{1}\otimes\sigma_{z}\otimes\mathbb{1} + \frac{\omega_L^{(2)}}{2}\mathbb{1}\otimes\mathbb{1}\otimes\sigma_{z} \\
%    &+ \frac{A_{\|}^{(1)}}{2}Z_{e}\otimes\sigma_{z}\otimes\mathbb{1} + \frac{A_{\perp}^{(1)}}{2}Z_{e}\otimes\sigma_{x}\otimes\mathbb{1}\\
%    &+ \frac{A_{\|}^{(2)}}{2}Z_{e}\otimes\mathbb{1}\otimes\sigma_{z} + \frac{A_{\perp}^{(2)}}{2}Z_{e}\otimes\mathbb{1}\otimes\sigma_{x}\\
%    =& \sum_{j=0}^{1}\sigma_{jj}\otimes \left(H_j^{(1)}\otimes\mathbb{1} + \mathbb{1}\otimes H_{j}^{(2)}\right)\, ,
%\end{split}
%\end{equation}
Neglecting direct inter-nuclear interactions, the Hamiltonian for $L$ nuclear spins is given by 
\begin{equation}
\begin{split}
    H =& \sum_{j=0}^{1}\sigma_{jj}\otimes\Big(H_j^{(1)}\otimes\mathbb{1}_{2^{L-1}} + \mathbb{1}\otimes H_{j}^{(2)}\otimes\mathbb{1}_{2^{L-2}}\\
    +& ... + \mathbb{1}_{2^{l-1}}\otimes H_{j}^{(l)}\otimes\mathbb{1}_{2^{L-l}} +... +\mathbb{1}_{2^{L-1}}\otimes H_{j}^{(L)}\Big)\,.
\end{split}
\end{equation}
where 
\begin{equation}
    H_j^{(l)} = \frac{\omega_{L}^{(l)} + s_{j}A_{\|}^{(j)}}{2}\sigma_{z}^{(l)} + \frac{s_{j}A_{\perp}^{(l)}}{2}\sigma_{x}^{(l)}\, ,
\end{equation}
where $\sigma_x^{(l)}$ and $\sigma_{z}^{(l)}$ are the Pauli operators acting only on the $l$-th nuclear spin (the identity operator acts on all the other spins). Notice that the Larmor frequency, $\omega_L^{(l)}$, is different for different types of nuclear spins. Recall that for $^{13}\rm C$ the gyromagnetic ratio is $\gamma_{^{13}\rm C} = 2\pi\cdot10.7084\, \rm MHz/T$, and for $^{29}\rm Si$ the ratio is $\gamma_{^{29}\rm Si} = -2\pi\cdot8.465\, \rm MHz/T$. 
The above secular form of the Hamiltonian make it apparent that the terms in the parentheses commute with each other, i.e., $[\mathbb{1}\otimes H_j^{(k)}\otimes\mathbb{1},\mathbb{1}\otimes H_j^{(l)}\otimes\mathbb{1}] = 0$, where $\mathbb{1}$ has the appropriate dimensions in each position. As a result of this feature, in the case of a single electronic spin coupled to multiple nuclear spins, the $\pi$-pulse sequences generate an evolution operator which is a sum of terms, namely 
\begin{equation}
    \label{eq:evolution_operator_multiple_nuclear_spins}
    U=\sum_{j=0}^{1}\sigma_{jj}\otimes_{l=1}^{L}R_{\mathbf{n}_j}^{(l)}\big(\phi_j^{(l)}(N)\big)\,,
\end{equation}
with $L$ denoting the total number of nuclear spins and $\phi_j^{(l)}(N)$ being the total angle of rotation after $N$ iterations. From now on, we will denote the total angle or rotation simply as $\phi_j^{(l)}$. The last equation reveals that the evolution operator of the total system is defined by the evolution of each nuclear spin and the electron. This feature enables us to derive analytical expressions for the average of the one-tangles across varying numbers of nuclear spins.\\
\indent In Ref.~\cite{Takou2023}, it was shown that the one-tangle of a single nuclear spin, when partitioned from the remaining electron-nuclear register, is given by 
\begin{equation}
    \label{eq:electron_nuclear_entangling_power}
    \epsilon_{p|q}^{\rm nuclear} = 1-G_{1}^{(q)}\,,
\end{equation}
where we omit the $2/9$ overall factor for simplicity, as we did in Eq.~\eqref{eq:entangling_power}, and $G_1^{(q)}$ denotes the first Makhlin invariant for the $q$-th nuclear spin. Equation~\eqref{eq:electron_nuclear_entangling_power} holds for $n\geq3$ qubits, while for $n=2$ qubits the average of the one-tangles is the same as the two-qubit entangling power given in Eq.~\eqref{eq:entangling_power}. It is worth noting that the one-tangle of a nuclear spin is solely affected by the factors governing its evolution, owing to the tensor product structure of the overall evolution operator $U$. Essentially, altering the partitioned nuclear spin within the register modifies Eq.~\eqref{eq:electron_nuclear_entangling_power} via its $G_1^{(q)}$, indicating a different level of entanglement between the partitioned spin and the remaining electron-nuclear register. Although Eq.~\eqref{eq:electron_nuclear_entangling_power} aligns with the two-qubit entangling power from  Eq.~\eqref{eq:entangling_power} in a disjointed view, the interpretations of these equations differ. More specifically, Eq.~\eqref{eq:entangling_power} tells us to what extent the two-qubit gate, created by the $\pi$-pulse sequences, can generate an electron-nuclear spin Bell state, while Eq.~\eqref{eq:electron_nuclear_entangling_power} describes correlations in the multi-spin register and, therefore, needs to respect the monogamy of entanglement \cite{Takou2023}.  
\subsection{\label{subsec:Single_Shot_entanglement}Multiple nuclear spins - Single-shot entanglement}
\indent In this subsection, we exploit the two-fold feature of DD sequences by entangling the electron selectively to a set of nuclear spin target(s) in a nuclear spin bath. In detail, we use DD sequences to couple the electron to a specific set while, simultaneously, decoupling it from the rest of the bath. We quantify the amount of target and unwanted entanglement using the one-tangles mentioned in Sec.~\ref{sec:Quantifying_Entanglement}. This approach of synchronous maximization of multiple one-tangles was first proposed in Ref.~\cite{Takou2023}.\\
\begin{figure}[b!]
    \includegraphics[width=1\linewidth,keepaspectratio]{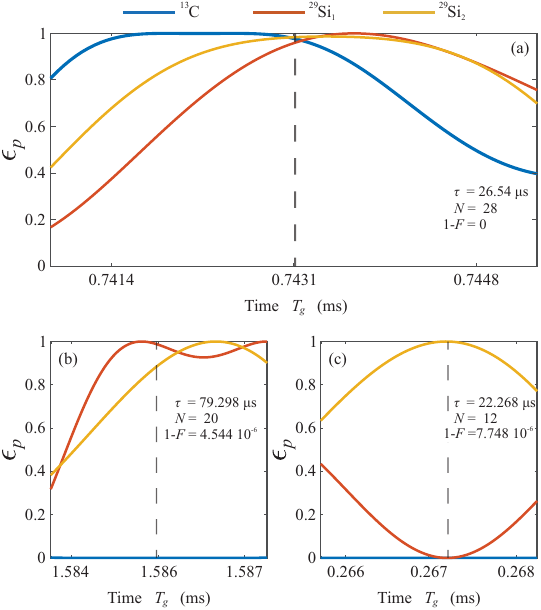}
    \caption{\label{fig:Ep_three_spins} Scaled one-tangles for three nuclear spins coupled to the electron. We present numerical results for the algorithm proposed in the main text using CPMG sequences for a set of (a) three spins, (b) two spins, and (c) one spin coupled to the electronic spin. Each graph depicts the entangling power between a single nuclear spin partitioned from the rest of the system as a function of $N\tau$, where $\tau$ is the unit time and $N$ the number of iterations. The horizontal axis in each panel corresponds to fixing $N$ to its optimal value and scanning $\tau$. The optimal values of $\tau$ and $N$ in each case are given in the graphs, while the vertical dashed line indicates the corresponding gate time $N\tau$ at the optimal point. The Larmor frequencies are $(\omega_L^{\rm C},\omega_L^{\rm Si}) =2\pi\cdot(88.8797,\,-70.2595)\,\rm kHz$, while the HF couplings are $(A_{||}^{\rm C},A_{\bot}^{\rm C})=2\pi\cdot(151.3741,105.0043\,)\, \rm kHz$, $(A_{||}^{\rm Si_1},A_{\bot}^{\rm Si_1})=2\pi\cdot(96.2445,180.9921\,)\, \rm kHz$, and $(A_{||}^{\rm Si_2},A_{\bot}^{\rm Si_2})=2\pi\cdot(122.1684, 123.7244\,)\, \rm kHz$. The HF values for the \textsuperscript{13}C and the \textsuperscript{29}Si are drawn from uniform distributions in the intervals $[10,200]\rm\, kHz$ and $[0.5, 200]\rm\, kHz$, respectively.  }
\end{figure}
\indent To illustrate the method, we consider a single defect and a total of three nuclear spins, \textsuperscript{13}C and \textsuperscript{29}Si. Each nuclear spin is characterized by its HF interaction parameters, and thus by its resonance times, approximated in Eq.~{(\ref{eq:resonance_times})}. There are two key parameters that we need to determine: the duration $\tau$ of one sequence unit and the number of iterations $N$. Our goal is to choose the unit time $\tau$ and the number of iterations $N$ such that we entangle a desired number of nuclear spins to the electron; in the example that follows this number can be 1, 2, or 3. For simplicity, we present the algorithm only for CPMG, but all the steps remain the same for UDD sequences too. As a first step, we calculate the resonance times $\tau_k^{(l)}$ for every nuclear spin, $l\in\{1,2,3\}$. To do so, we use Eq.~\eqref{eq:resonance_times} to reach the vicinity of the resonance, and then we find the precise value of $\tau_k^{(l)}$ numerically since we know that at this value the dot product of the axes of rotation, for the respective spins, is equal to $-1$, i.e., $\mathbf{n}_0\cdot\mathbf{n}_1 = -1$. As mentioned above, perfect entanglement can be achieved as long as $\mathbf{n}_0\cdot\mathbf{n}_1\leq 0$, so we allow the unit time to take values around the resonance times,  $\tau \in[\tau_k^{(l)}-\delta\tau,\,\tau_k^{(l)}+\delta\tau]$. Next, we find the one-tangle, $\epsilon_p^{(l)}(\tau, N)$, for every nuclear spin from Eq.~\eqref{eq:electron_nuclear_entangling_power}, which is a function of the unit time $\tau$ chosen in the above intervals, and the number of repetitions $N$. We also set a maximum gate time $T_g$ so that $N\tau\leq T_g$. All the information we need to choose $\tau$ and  $N$ is given by $\epsilon_p^{(l)}(\tau, N)$. For instance, if we want to entangle all three spins with the electron we must choose a pair of $(\tau, N)$ such that $\epsilon_p^{(l)}$ is above a threshold $\epsilon_p^{\rm th}$ and close to unity for $l=1,2,3$. Another case would be to create a register of two nuclear spins, known as target spins, and decouple the register from the third nuclear spin. This can be achieved by choosing a pair of $(\tau, N)$ such that two one-tangle values, $\epsilon_p^{(l)}(\tau, N)$, are maximized and larger than $\epsilon_p^{\rm th}$ while the third one is minimized and smaller than $\epsilon_p^{\rm th}$. \\
\indent To make the above example more concrete, in Fig.~\ref{fig:Ep_three_spins} we present a numerical application of the proposed recipe. In Fig.~\ref{fig:Ep_three_spins} we show that for a silicon vacancy, with quantum numbers $(s_0,s_1) = (1/2,3/2)$, and three nuclear spins (one \textsuperscript{13}C, two \textsuperscript{29}Si), we can selectively entangle nuclear spins with the electron, or decouple them from it, simply by using different combinations of $(\tau,N)$. Figure~{\ref{fig:Ep_three_spins}(a)} shows that a register of three spins and the electron can be created by setting $(\tau,\, N) = (26.54\,\mu{\rm s},\,28)$ for a CPMG sequence, while the size of the register is reduced to two nuclear spins in Fig.~{\ref{fig:Ep_three_spins}(b)} by choosing $(\tau,\,N)=(79.298\,\mu{\rm s}, 20)$, and to one nuclear spin in Fig.~{\ref{fig:Ep_three_spins}(c)} for $(\tau,\,N)=(22.268\,\mu{\rm s}, 12)$. Note that in Fig.~\ref{fig:Ep_three_spins}(a), we simultaneously optimize all three one-tangles, while in the other two panels, we minimize the unwanted one-tangles but do not attempt to maximize the wanted ones. This combination of features, being able to change the size of the register while effectively reducing the cross-talk from the bath, is essential for creating a quantum register. Moreover, the total gate time needed is always shorter than the nuclear spin coherence time $T_2^*$ which ranges from 3 to 17 ms \cite{Bradley2021}. Also, the reduction of the cross-talk is of significant importance because, as we will show in the next section, it affects the gate fidelity. In the next section, we extend the algorithm by including more DD sequences and a range of register sizes and bath qubit assignments.\\
\section{Fidelity dependence on register and bath size \label{sec:Results}}
\indent In this section, we quantify the effect of the register size and the spin bath on the gate fidelity in the case of silicon monovacancies and divacancies coupled to a nuclear bath of \textsuperscript{13}C and \textsuperscript{29}Si nuclear spins. We begin by introducing a way to optimize the gate fidelity. We then implement this method for a monovacancy coupled to multiple nuclear spins, and, finally, we examine the case of a mixed bath where the nuclear spins have the same sign in gyromagnetic ratio.\\
\subsection{Fidelity optimization}
\begin{figure*}\includegraphics[width=1\textwidth,keepaspectratio]{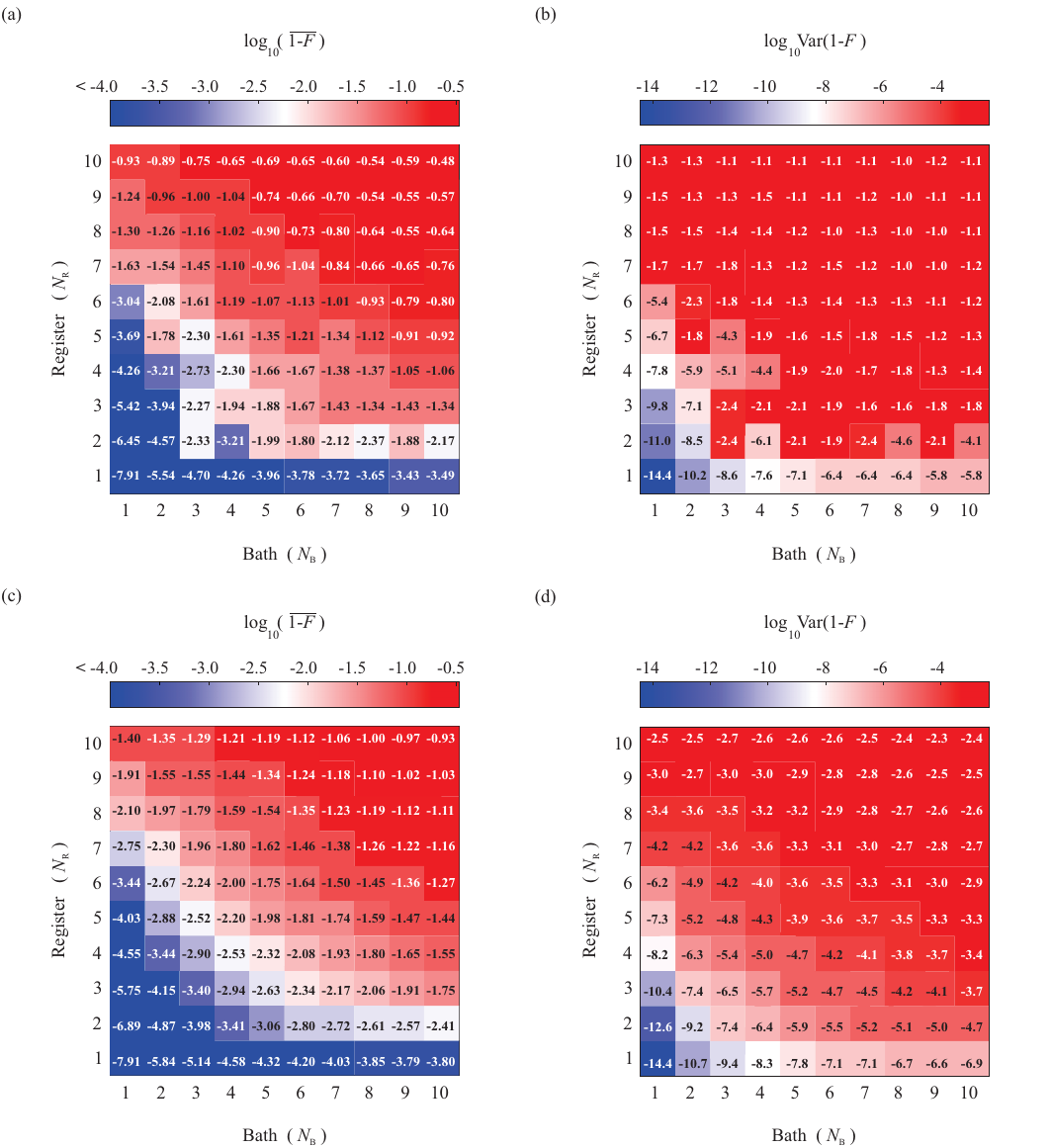}
    \caption{ \label{fig:Fidelity_results}Gate infidelity $(1-F)$ versus the size of the bath (unwanted spins) and the size of the register (wanted or target spins) before (a,b) and after (c,d) optimization over single-qubit rotations on the electronic spin. In (a) and (c) we plot the decimal logarithm of the statistical average of infidelity, i.e., $\log_{10}\left(\overline{1-F}\right)$, while in (b) and (d) we plot the variance of the infidelity on the same logarithmic scale $\log_{10}{\rm Var}\left(1-F\right)$. The numbers in each tile give the value of $\log_{10}(\overline{1-F})$ (a,c) and $\log_{10}{\rm Var}(1-F)$ (b,d). In the above simulations, we considered SiC samples with a silicon monovacancy with total spin $S = 3/2$, and the electron qubit is defined by $(s_0,s_1)=(1/2,3/2)$. The HF values for the \textsuperscript{13}C and the \textsuperscript{29}Si were drawn from uniform distributions in the intervals $2\pi[10,200]\,\rm kHz$ and $2\pi[0.5, 200]\,\rm kHz$, respectively. The entangling power threshold was set to $\epsilon_p^{\rm th} = 0.85$, the magnetic field is $B = 83\rm G$ leading to Larmor frequencies $ (\omega_L^{C},\omega_L^{Si}) = 2\pi( 88.8797,\, -70.2595)\, \rm kHz$, and the relative abundance was kept constant for all the cases, i.e., $\frac{{}^{29}\rm{Si}}{^{13}\rm{C}} = \frac{4.7\%}{1.1\%} = 4.27$. Every data point on the heatmaps was obtained from 200 realizations, and for each trajectory we kept the combination $(\tau,N)$ that resulted in the lowest maximum unwanted one-tangle.}
\end{figure*}
\indent Let's consider a scenario with a total of $L$ nuclear spins, where $K$ of them are the target nuclei exhibiting maximal one-tangles. The unwanted nuclei, $ L - K$, impact the target gate because they generally possess some residual level of entanglement with the electron. Consequently, truncating the evolution operator to the target subspace results in a non-unitary operation. Ref.~\cite{Takou2023} demonstrated how to circumvent this issue using the Kraus operator representation of the partial trace channel. This approach allows us to directly manipulate the total evolution operator without specifying an initial state for the system. We use the analytical expression for the gate fidelity within the target subspace by employing the operator-sum representation \cite{Takou2023}. As the target gate we consider the evolution operator
\begin{equation}
    \label{eq:target_gate_U0}
    U_0 = \sum_{j=0,1}\sigma_{jj}\otimes_{k = 1}^{K}R_{\mathbf{n}_j^{(k)}}\big(\phi_j^{(k)}\big)\,,
\end{equation}
that we design in the ideal case of no unwanted spins. This target gate is ignorant of the presence of unwanted spins, i.e., it is defined in the truncated space of the electron and the $K$ target nuclear spins. The total evolution operator, given in Eq.~{(\ref{eq:evolution_operator_multiple_nuclear_spins})}, can be re-written as
\begin{equation}
    \label{eq:total_evol_oper_permutated}
    U = \sum_{j = 0,1}\sigma_{jj}\otimes_{k=1}^KR_{\mathbf{n}_j^{(k)}}\big(\phi_j^{(k)}\big)\otimes_{l=1}^{L-K}R_{\mathbf{n}_j^{(K+l)}}\big(\phi_j^{(K+l)}\big)\,,
\end{equation}
where, for convenience, we permuted the spins such that the register spins appear first in the tensor product with the electron spin projector and the unwanted spins appear in the last positions.\\
\indent In the context of the operator-sum representation \cite{Nielsen2010}, the fidelity of a quantum operator has the following form
\begin{equation}
    \label{eq:Fidelity_general_formula}
    F = \frac{1}{d(d+1)}\sum_j {\rm tr}[(U_0^{\dagger}E_j)^{\dagger}U_0^{\dagger}E_j] + |{\rm tr}[U_0^\dagger E_j]|^2\, ,
\end{equation}
where $d = 2^{K+1}$ represents the dimension of the target subspace, which includes the electron and $K$ target spins. The index $j$ iterates through the environment's $2^{L-K}$ complete computational basis states, the Kraus operators $E_j$ describe the quantum channel denoted by $\mathcal{E}(\rho) = \sum_j E_j\rho E_j^\dagger$ and they satisfy the completeness relation $\sum_j E_j^\dagger E_j = 1$. The Kraus operators are given by \cite{Takou2023}
\begin{equation}
    \label{eq:Kraus_operators}
    E_i = \sum_j c_j^{(i)} p_j^{(i)} \sigma_{jj} \otimes_{k=1}^KR_{\mathbf{n}_j^{(k)}}\big(\phi^{(k)}_j\big)\, ,
\end{equation}
where $c_j^{(i)}\equiv \prod_{m=m_1}^{m_M}[\cos(\phi_j^{(m)}/2) - in_{z,j}^{(m)}\sin(\phi_j^{(m)}/2)]$ and $p_j^{(i)}\equiv \prod_{\xi=\xi_1}^{\xi_{L-K-M}}[-(in_{x,j}^{(\xi)} + in_{y,j}^{(\xi)})\sin(\phi_j^{(\xi)}/2)]$ while $\{n_x,n_y,n_z\}$ correspond to the rotation axis components of each nuclear spin. Also, $M$ counts the number of the environment spins in the $\ket{0}$ state, while the other $K-L-M$ are in the $\ket{1}$ state. In the case where $M=0$ so that all the unwanted spins are in the $\ket{1}$ state, it holds that $c_j^{(i)} = 1$, and when $M=L-K$, all the unwanted spins are in the $\ket{0}$ state, in which case $p_j^{(i)} = 1$.\\
\indent Substituting Eqs.~\eqref{eq:target_gate_U0} and \eqref{eq:Kraus_operators} into Eq.~\eqref{eq:Fidelity_general_formula}, the gate fidelity becomes
\begin{equation}
    \label{eq:fidelity_not_optimized}
    F = \frac{1}{2^{K+1}+1}\Big( 1 + 2^{K-1}\sum_k\bigg\lvert\sum_{j=0,1}c_j^{(k)}p_j^{(k)}\bigg\rvert^2\Big)\, .
\end{equation}
Obviously, in the absence of unwanted nuclear spins the fidelity is unity. The data in Fig.~\ref{fig:Ep_three_spins} reveal the direct connection between the one-tangles and the fidelity of the target gate. We also find that the fidelity of the target gate is mostly affected by the largest one-tangle of the unwanted spins, which agrees with Ref.~\cite{Takou2023}. This last observation is especially important since it identifies the parameter we should minimize in order to maximize the target gate's fidelity. As mentioned in Ref.~\cite{Takou2023}, the gate error can exhibit sudden spikes that reach significantly high levels even at low one-tangle values. This behavior arises because the presence of unwanted spins can disrupt the expected ideal evolution of the isolated target system described by Eq.~\eqref{eq:target_gate_U0}. However, despite this deviation, the resulting gate operation might exhibit a greater overlap with other entangling gate operations that are equivalent to Eq.~\eqref{eq:target_gate_U0} up to local gates. To account for this, we can generalize the target gate to include an arbitrary single-qubit rotation on the electronic spin, $R_{\mathbf{n}^e}(\theta)$, and then optimize over the axis $\mathbf{n}^e$ and angle $\theta$ of the rotation. The target gate thus becomes, $\tilde{U}_0 = R_{\mathbf{n}^e}(\theta)U_0$, leading to the fidelity (see Appendix~\ref{App:Gate_fidelity_optimization} for the proof)
\begin{equation}
    \label{eq:fidelity_optimized}
    \begin{split}
    F_{\rm opt} =& \frac{1}{2^{K+1}+1}\Big( 1 + 2^{K-1} \times \\&\sum_k\bigg\lvert\sum_{j=0,1}c_j^{(k)}p_j^{(k)}\bigg\{\cos\left(\frac{\theta}{2}\right)\\
  &\qquad\qquad\qquad\quad\,+i(-1)^j\sin\left(\frac{\theta}{2}\right) n_z^{e}\bigg\}\bigg\rvert^2\Big)\, .
    \end{split}
\end{equation}
In principle, we can optimize the above relation over the range $\theta\in[0,2\pi)$ for the rotation angle, and $n_z^e\in[-1,1]$ for the $z$ component of the rotation axis. However, the maximum fidelity can be achieved only when $|n_z^e|=1$. This can be understood by simply noticing that the fidelity in Eq.~{(\ref{eq:fidelity_optimized})} is maximized when the second term in the parentheses becomes $2^{K-1}\cdot2^2$, and this cannot happen when the factors in the summation are multiplied with a complex number which has norm smaller than unity. In light of this, the rotation axis must be set to $\mathbf{n}^e = \mathbf{z}$, and then we can optimize over the rotation angle $\theta\in[0,4\pi)$.\\
\subsection{Average fidelity as a function of register and bath size}
\begin{table}
\centering % instead of \begin{center}
\caption{\label{tab:fitting_parameters} Fitting parameters for the fit function $\log_{10}(\overline{1-F_{N_R}}) = a_{N_R}e^{b_{N_R}N_B} + c_{N_R}e^{d_{N_R}N_B}$.}
\begin{tabular}{SSSSS} 
\toprule\toprule
    {$N_R$} & {$a$} & {$b$} & {$c$} & {$d$} \\ \midrule
    1  & -7.594 & -0.876 &  -4.857 & -0.026 \\
    2  & -7.384  & -0.712 & -3.365  & -0.032 \\
    3  & -5.372  & -0.833 & -3.665 & -0.073 \\
    4  & -3.556 & -0.873 & -3.306  & -0.076 \\ 
    5  & -4.041  & -1.075 & -2.835   & -0.071 \\
    6  & -2.577  & -0.641 & -2.196  & -0.054 \\
    7  & -1.530  & -0.476 & -1.895  & -0.051 \\
    8  & -2.322  & -0.089 & -8.5$\cdot 10^{-5}$  & 0.758\\ 
    9  & -3.4$\cdot 10^6$  & -16.56 & -1.798  & -0.059 \\
    10 & -1.476   & -0.046 & 0   & 0 \\
 \bottomrule \bottomrule
\end{tabular}
\end{table}
\begin{figure}
    \includegraphics[width=\linewidth,keepaspectratio]{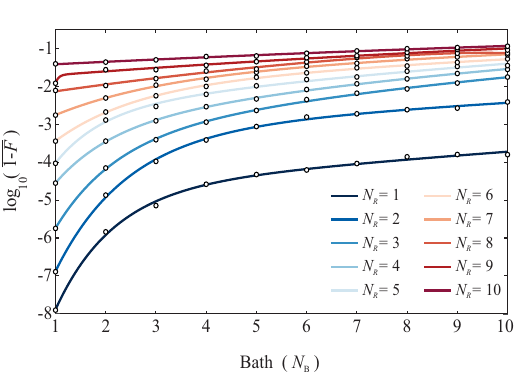}
    \caption{ \label{fig:fig_fit} Statistical average of the infidelity (circles) and fits (solid lines) versus the spin bath size. The lines are fits of the function $\log_{10}(\overline{1-F_{N_R}}) = a_{N_R}e^{b_{N_R}N_B} + c_{N_R}e^{d_{N_R}N_B}$ to the collected data (circles). Each line refers to a different register size $N_R$, and the respective fitting parameters $a,b,c,d$ are given in Table~\ref{tab:fitting_parameters}.}
\end{figure}
\begin{figure*}\includegraphics[width=\textwidth,keepaspectratio]{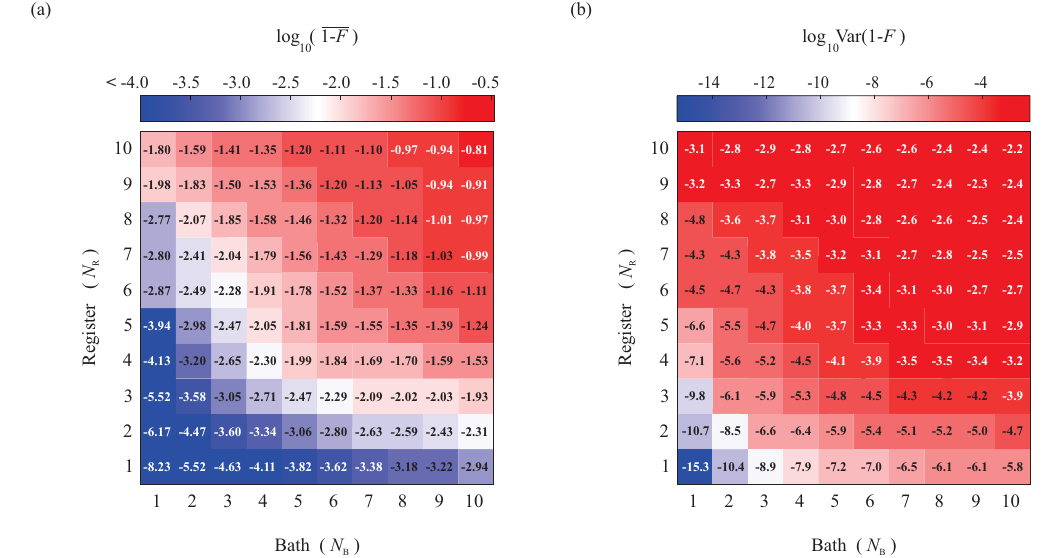}
    \caption{ \label{fig:Fidelity_results_Divacancy}Gate infidelity $(1-F)$ versus the size of the bath (unwanted spins) and the size of the register (wanted or target spins) upon optimization over single-qubit rotations on the electronic spin. In (a) we plot the decimal logarithm of the statistical average of infidelity, i.e., $\log_{10}\left(\overline{1-F}\right)$, while in (b) we plot the variance of the infidelity on the same logarithmic scale $\log_{10}{\rm Var}\left(1-F\right)$. The numbers in each tile give the value of $\log_{10}(\overline{1-F})$ and $\log_{10}{\rm Var}(1-F)$, respectively. In the above simulations, we considered SiC samples with a silicon divacancy with total spin $S = 1$, and the electron qubit is defined by $(s_0,s_1)=(0,-1)$. The HF values for the \textsuperscript{13}C and \textsuperscript{29}Si nuclei were drawn from uniform distributions in the intervals $2\pi[10,200]\,\rm kHz$ and $2\pi[0.5, 200]\,\rm kHz$, respectively. The entangling power threshold was set to $\epsilon_p^{\rm th} = 0.85$, the magnetic field is $B = 584\rm G$ leading to Larmor frequencies $ (\omega_L^{C},\omega_L^{Si}) = 2\pi( 625.37,\, -494.35)\, \rm kHz$, and the relative abundance was kept constant for all the cases, i.e., $\frac{{}^{29}\rm{Si}}{^{13}\rm{C}} = \frac{4.7\%}{1.1\%} = 4.27$. Every data point on the heatmaps was obtained from 100 realizations, and for each trajectory we kept the combination $(\tau,N)$ that resulted in the lowest maximum unwanted one-tangle.}
\end{figure*}
\indent In the following simulations, we presume an electron spin with $ S = 3/2$ with the qubit defined using the $(s_0,s_1) = (1/2,3/2)$ spin states, which correspond to a silicon monovacancy defect in SiC \cite{Widmann2014, Nagy2019, Babin2021}. The nuclear Larmor frequencies are specified as $ (\omega_L^{\rm C},\omega_L^{\rm Si}) = 2\pi( 88.8797,\, -70.2595)\, \rm kHz$ for $ ^{13}\rm C $ and $ ^{29}\rm Si $ nuclei, respectively. In the absence of experimental data, we draw the HF values from uniform distributions in the range $ 2\pi[10,200]\,\rm kHz $ and $ 2\pi[0.5, 200]\,\rm kHz $ for $ ^{13}\rm C $ and $ ^{29}\rm Si $ nuclei, respectively. These nuclei are weakly coupled since the HF parameters are smaller than $ 1/T_2^* $, which is typically a few hundred kHz for NV centers \cite{Maze2012, Liu2012} or, in general, when $ A_{\|},A_{\bot}\ll 1 \,\rm MHz $ \cite{Ghosh2021} (approximately 1 MHz is also the electron linewidth for the neutral divacancy in SiC \cite{Bourassa2020}). Moreover, to ensure the distinctness of spins chosen through random generation, we impose a condition on the discrepancy of HF values; we stipulate that at least one HF value ($A_{\|}$ or $A_{\perp}$) should deviate by at least $ 2\pi\cdot10\, \rm kHz $ from the others. This condition is chosen to ensure the generation of sufficient nuclei within the HF range with distinct HF values in order to simulate a realistic silicon vacancy in a SiC sample. Our goal is to examine the dependence of the fidelity on several register-bath size combinations. To do so, we use the algorithm described in Sec.~{\ref{subsec:Single_Shot_entanglement}}, aiming to maximize the one-tangles of the target spins while minimizing those of the unwanted spins. Contrary to the simple example shown in Fig.~\ref{fig:Ep_three_spins}, now will we explore the capabilities of five resonances ($k \in \{1,2,3,4,5\}$) of the CPMG, UDD\textsubscript{3}, and UDD\textsubscript{4} pulse sequences.\\
\indent  In Fig.~\ref{fig:Fidelity_results}, we show results from 200 realizations for the case where we have a silicon vacancy defect coupled to $N_{\rm R} + N_{\rm B}$ total nuclear spins, where $N_{\rm R}$ and $N_{\rm B}$ are the number of nuclear spins in the register and the bath, respectively. In the left column of Fig.~\ref{fig:Fidelity_results} we present the statistical average of the infidelity, $\overline{1-F}$, and in the right column its variance, ${\rm Var}(1-F)$, both on a logarithmic scale. For instance, a square data point depicts the statistical average (or variance) of infidelity for a register size $N_{\rm R}$ and bath size $N_{\rm B}$ over 200 realizations. The top set of heatmaps shows the fidelity and the variance obtained from Eq.~{(\ref{eq:fidelity_not_optimized})}, while the lower set shows the same quantities after the optimization over single-qubit rotations on the electron spin, see Eq.~{(\ref{eq:fidelity_optimized})}. Comparing the two rows, one can see that in most cases both the infidelity and its variance become smaller, meaning that the optimization leads to an increase in average fidelity, and the fidelities are concentrated around the average. The highest gain occurred for $(N_B,N_R)=(3,2)$ where the infidelity dropped by $1.6$ orders of magnitude, while the variance dropped by five orders of magnitude. Now let us focus on the lower set of heatmaps. As expected, in Fig.~{\ref{fig:Fidelity_results}(c)}, the best fidelity is obtained in the case of a single target and a single unwanted spin. Also, for small register size, $N_R\leq 5$, there is an exponential decrease in fidelity as the bath size $N_B$ increases. Moreover, Fig.~{\ref{fig:Fidelity_results}(c)} reveals a clear deterioration of fidelity not only with the bath size but also with the register size. This can be explained by considering that the generation of entanglement links between the electron and the register spins, while simultaneously decoupling unwanted spins using DD pulses, becomes very challenging when the register size gets larger. The same happens with the variance, where we see that it grows with the total number of spins, leading to an increasing spread in the fidelity distribution, and thus the uncertainty for the achievable fidelity. The heatmaps in Fig.~\ref{fig:Fidelity_results} are intended to serve as a guide for experimentalists working with the silicon vacancy in SiC and considering the benefits of isotopically purifying their samples.\\
\indent As an extension of the above results, we fit the statistical average of the infidelity shown in Fig.~{\ref{fig:Fidelity_results}(c)} using a double exponential formula. More precisely, we use $\log_{10}(\overline{1-F_{N_R}}) = a_{N_R}e^{b_{N_R}N_B} + c_{N_R}e^{d_{N_R}N_B}$ to fit across each row of Fig.~{\ref{fig:Fidelity_results}(c)}, where the index $N_R\in[1,10]$ denotes the size of the register and $N_B\in[1,10]$ denotes the size of the bath. The fit parameters can be found in Table~\ref{tab:fitting_parameters} while the data and the fit functions are shown together in Fig.~\ref{fig:fig_fit}. These functions can be used to estimate the average fidelity in the presence of more unwanted spins.\\
\indent Next, we consider the case of a silicon divacancy defect in SiC with electron spin $S = 1$ where the qubit is defined using the $(s_0,s_1) = (0,-1)$ spin states. Similarly to what we showed for the monovacancy center above, in Fig.~\ref{fig:Fidelity_results_Divacancy} we show the statistical average of the infidelity and its variance over 100 realizations in a much stronger magnetic field, $B = 584\, \rm G$, taken from Ref.~\cite{Bourassa2020}. Comparing Fig.~\ref{fig:Fidelity_results_Divacancy}(a) and Fig.~\ref{fig:Fidelity_results}(c), we observe that despite the difference in the spin states used to define the electronic spin qubit and the strength of the applied magnetic field, the fidelity deteriorates in the same fashion with the size of the register, while it decreases faster with the number of the unwanted spins. However, there are cases where the divacancy and a strong magnetic field lead to higher average fidelity; compare for instance the $N_B=1$ columns in Fig.~\ref{fig:Fidelity_results_Divacancy}(a) and Fig.~\ref{fig:Fidelity_results}(c). The last two features can be attributed to the fact that the combination of total spin $S=1$ and strong magnetic field lead to resonance times with lower dispersion, hence making it easier for DD pulses to create large registers but more difficult to decouple selected nuclear spins from a large bath. Also, low dispersion in resonance times can affect our ability to create a desired entangled state in the presence of a given number of unwanted spins (see Appendix~\ref{App:Success_Probability}). 
\begin{figure}
    \includegraphics[width=1\linewidth,keepaspectratio]{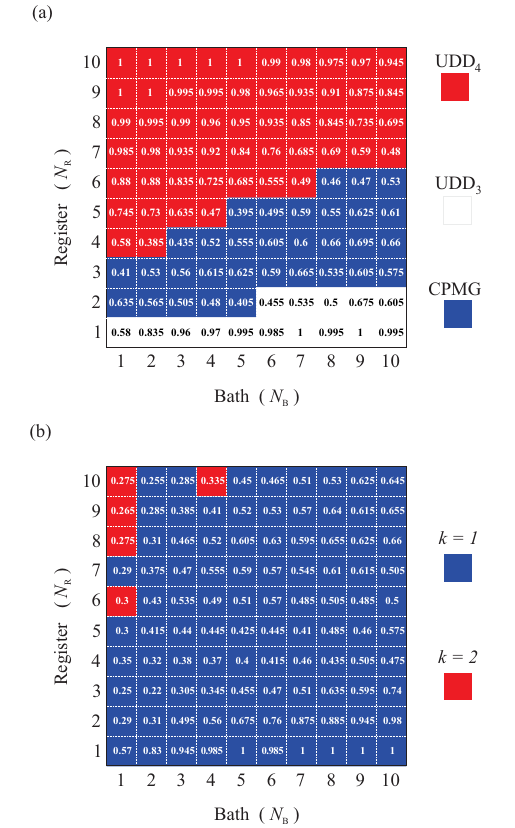}
    \caption{ \label{fig:Most_used_sequences} Best pulse sequence and resonance order versus the size of the bath (unwanted spins) and the size of the register (wanted or target spins). (a) Heatmap depicting the most used (or most probable) DD sequence among CPMG (blue), UDD\textsubscript{3} (white), and UDD\textsubscript{4} (red). (b) Heatmap showing the most used (or most probable) resonance order ($k=1$ blue, $k=2$ red) used for $\tau_k$. The number in each tile refers to the occurrence probability. In the above simulations, we considered SiC samples with a silicon monovacancy with total spin $S = 3/2$ where the electron qubit is defined by $(s_0,s_1)=(1/2,3/2)$. The HF values for the \textsuperscript{13}C and \textsuperscript{29}Si spins were drawn from uniform distributions in the intervals $2\pi[10,200]\,\rm kHz$ and $2\pi[0.5, 200]\,\rm kHz$, respectively. The entangling power threshold was set to $\epsilon_p^{\rm th} = 0.85$, the magnetic field was $B = 83\rm G$ leading to Larmor frequencies $ (\omega_L^{C},\omega_L^{Si}) = 2\pi( 88.8797,\, -70.2595)\, \rm kHz$, and the relative abundance was kept constant for all the cases, i.e., $\frac{{}^{29}\rm{Si}}{^{13}\rm{C}} = \frac{4.7\%}{1.1\%} = 4.27$. Every data point on the heatmaps was obtained from 200 realizations, and for each trajectory we kept the combination $(\tau,N)$ that resulted in the lowest maximum unwanted one-tangle. }
\end{figure}
\subsection{Performance of DD sequences}
\indent As a next step, we investigate the suitability of the different DD sequences and the resonance orders versus the total number of nuclear spins. To that end, in Fig.~{\ref{fig:Most_used_sequences}(a)} we present the probability of a given DD sequence to achieve the highest fidelity. The probability here refers to the number of times that a specific sequence was used in 200 independent realizations. Figure~{\ref{fig:Most_used_sequences}(a)} shows that UDD\textsubscript{4} performs much better for a higher number of qubits and specifically for $N_R\geq5$ and is the most effective sequence, with probability close to unity, for $N_R\geq 8$. In the same heatmap, we see that UDD\textsubscript{3} is the preferred DD sequence to use for register sizes $N_R=1$, 2, but for large baths, $N_B\geq 6$. Also, CPMG is found to give the best gate fidelity for registers of intermediate size, namely for $N_R\in[2,6]$ and various bath sizes. However, the occurrence probability is not as high, meaning other sequences perform comparably. In Fig.~{\ref{fig:Most_used_sequences}(b)}, we show the probability of a given resonance order to achieve the highest fidelity independently of the DD sequence used. We observe that the first resonance, $k=1$ (blue tiles), is highly preferred in all sequences, however, the occurrence probability is not considerably high, meaning that the DD sequences perform their best when they have access to a large variety of resonances.\\
\begin{figure}
    \includegraphics[width=1\linewidth,keepaspectratio]{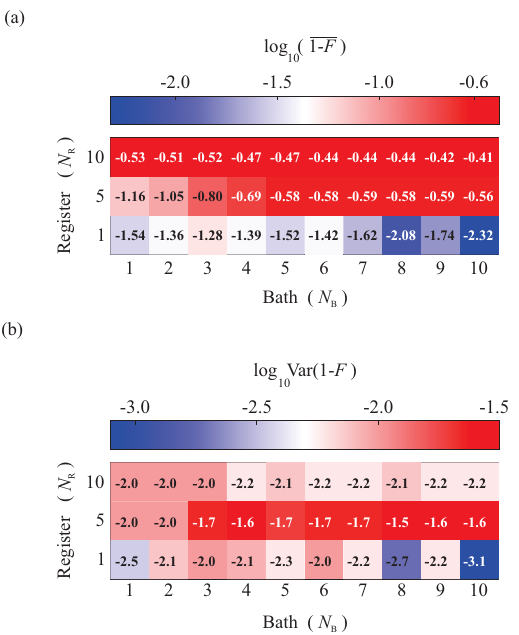}
    \caption{ \label{fig:Positive_gSi} Effect of changing the sign of a nuclear gyromagnetic ratio. Gate infidelity $(1-F)$ versus the size of the bath (unwanted spins) and the size of the register (wanted or target spins) upon optimization over single-qubit rotations on the electronic spin. (a) The decimal logarithm of the statistical average of the infidelity, i.e., $\log_{10}\left(\overline{1-F}\right)$. (b) The variance of the infidelity on the same logarithmic scale $\log_{10}{\rm Var}\left(1-F\right)$. The number on each tile gives the value of $\log_{10}(\overline{1-F})$ (a) and $\log_{10}{\rm Var}(1-F)$ (b). In the above simulations, we considered fictitious SiC samples with a silicon monovacancy with total spin $S = 3/2$, and the electron qubit is defined by $(s_0,s_1)=(1/2,3/2)$. The HF values for the \textsuperscript{13}C and the \textsuperscript{29}Si were drawn from uniform distributions in the intervals $2\pi[10,200]\,\rm kHz$ and $2\pi[0.5, 200]\,\rm kHz$, respectively. The magnetic field was $B = 83\rm G$ leading to Larmor frequencies $ (\omega_L^{C},\omega_L^{Si}) = 2\pi( 88.8797,\, 70.2595)\, \rm kHz$, and the relative abundance was kept constant for all the cases, i.e., $\frac{{}^{29}\rm{Si}}{^{13}\rm{C}} = \frac{4.7\%}{1.1\%} = 4.27$. Every data point on the heatmaps was obtained from 100 realizations, and for each trajectory we kept the combination $(\tau,N)$ that resulted in the lowest maximum unwanted one-tangle.}
\end{figure}
\subsection{Examination of positive gyromagnetic ratio}
\indent Finally, we examine the fictitious scenario where the \textsuperscript{29}Si nuclear spin has positive gyromagnetic ratio, i.e., $\gamma_{^{29}\rm Si} = +2\pi\cdot8.465\, \rm MHz/T $. We are interested in this scenario because it will help us conclude whether a mixed bath where the gyromagnetic ratios of the nuclei have opposite signs is preferable over the case where they have the same signs. In Fig.~\ref{fig:Positive_gSi} we plot the statistical average of the infidelity and the variance over 100 realizations upon optimization over single-qubit rotations on the electronic spin. We use the same algorithm as in Fig.~\ref{fig:Fidelity_results}, with the only difference being the sign of $\gamma_{^{29}\rm Si}$. Comparing Fig.~{\ref{fig:Fidelity_results}(c)-(d)} with Fig.~{\ref{fig:Positive_gSi}(a)-(b)}, one can easily see that a mixed nuclear bath with opposite signs in gyromagnetic ratios achieves at least two orders of magnitude lower infidelity for $N_R\leq 5$ and $N_B\leq 4$. Also, the performance of opposite signs remains better even in the challenging case of $N_R = 10$ nuclear spins, while the variance is approximately the same. Furthermore, one would expect the two cases to perform approximately the same for $N_R = 1$, however, this is where the differences become most significant. This can be explained by looking at the resonance times: The resonance times are larger in the case of negative $\gamma_{^{29} \rm Si}$ due to the sign difference between the Larmor frequency and the HF values (see Eq.~\eqref{eq:resonance_times}). Large resonance times lead to more distinct resonances between different nuclear spins, translating to easier decoupling. Overall, the above evidence indicates that a mixed nuclear bath with different signs in the gyromagnetic ratio leads to higher fidelities. It has been previously observed that having two nuclear species is beneficial for coherence times \cite{Bourassa2020,Widmann2014}; here, we see that it is also beneficial for entangling gates.
\section{\label{Sec:Conclusion} Conclusions}
Nuclear spins play a crucial role in spin-based solid-state platforms for quantum memories. To fully leverage their potential for creating large-scale quantum networks, a comprehensive investigation of the ability to control nuclear spin registers in all the promising candidate platforms should be conducted. Here, we developed a method for conducting such a high-throughput sorting of host materials and defects. As a concrete example to showcase our approach, we explored the performance of dynamical decoupling sequences such as CPMG, UDD\textsubscript{3}, and UDD\textsubscript{4} in controlling the nuclear spins surrounding a silicon monovacancy or divacancy defect in SiC. In each case, we quantified the expected performance in terms of the gate infidelity, and found an exponential deterioration not only with the bath size but also the register size. Additionally, we showed that some sequences perform better on average than others, depending on the register and bath size.\\
\indent In the pursuit of advancing quantum memory technologies, our work lays a foundation for further exploration. Future research could focus on extending our high-throughput sorting approach to other promising defects and host materials, and on combining it with  high-throughput electronic structure calculations. Investigating novel control schemes and pulse sequences, beyond CMPG and UDD, could enhance the performance of quantum memories, enabling more efficient and reliable quantum information processing. Moreover, integrating experimental efforts with theoretical advancements will be essential for moving this field forward. By continuing to push the boundaries of quantum memory research, we can unlock new opportunities for the development of robust and scalable quantum networks.\\
\section*{Acknowledgments}
SEE acknowledges support from the National Science Foundation (grant no. 2137645).
EB also acknowledges support from the National Science Foundation (grant nos. 1847078 and 2137953).

\appendix

\section{\label{App_1}UDD sequences}

For UDD\textsubscript{n} sequences, the time intervals between the $\pi$ pulses are given by 
\begin{equation}
\label{App:eq:UDDn_time_intervals}
    q_{r} = \sin^{2}\left(\frac{\pi r}{2n + 2}\right) - \sin^{2}\left(\frac{\pi(r-1)}{2n+2}\right)\, ,
\end{equation}
where $r$ takes integer values between $1$ and $n+1$, since there are $n+1$ time intervals. It is easy to show that $q_{r} = q_{n+2 -r}$, leading to symmetric time intervals about the halfway point of the sequence. However, this symmetry does not necessarily translate to equal angles of rotation. For UDD\textsubscript{4}, $[q_1\tau -\pi -q_2\tau-\pi-q_3\tau-\pi-q_4\tau-\pi-q_5\tau]$, it holds that $\phi_0\neq\phi_1$.\\
\indent As mentioned in the main text, UDD\textsubscript{n} sequences with odd $n$ (i.e., odd number of $\pi$ pulses) need to be repeated twice to form a ``unit sequence". This stems from the fact that the electron must return to its initial state. So, the UDD\textsubscript{3} unit is given by $[q_1\tau/2-\pi -q_2\tau/2-\pi-q_3\tau/2-\pi-(q_4+q_1)\tau/2-\pi-q_2\tau/2-\pi -q_3\tau/2-\pi-q_4\tau/2]$, where we divided all $q_r$ by a factor of $2$ so that the total time adds up to $\tau$.

\section{\label{App:Makhlin_invariants}Makhlin invariants}
The evolution operator of an electron spin and a lone nuclear spin, given in Eq.~\eqref{eq:Propagator_single_pulse}, can be characterized using the Makhlin invariants \cite{Makhlin2002}, commonly denoted by $G_1\in \mathds{C}$ and $G_2\in\mathds{R}$. These invariants categorize all two-qubit operations into separate entangling categories, ensuring that gates with identical local invariants belong to the same entangling category. This characteristic arises because local operations do not affect the amount of entanglement between the two parties. Gates capable of generating maximal entanglement, known as perfect entanglers, include the {\footnotesize{CNOT}} and {\footnotesize{CZ}} gates, which are equivalent up to single-qubit rotations. These gates, and any other two-qubit gate locally equivalent to them, define the category of perfect entanglers with $G_1 = 0$ and $G_2 = 1$.\\
\indent For any given $\pi$-pulse sequence repeated $N$ times, the electron-nuclear evolution operator maintains the structure described in Eq.~\eqref{eq:Propagator_single_pulse}, with the only difference being that $\phi_{j}$ is now replaced by the total rotation angle $\phi_{j}(N)$. This specific form of the evolution operator enables us to derive analytical expressions for $G_1$ and $G_2$. Understanding these two parameters helps us identify conditions under which the driven electron-nuclear evolution achieves perfect entanglement. Assuming an arbitrary $\pi$-pulse sequence and using Eq.~\eqref{eq:Propagator_single_pulse}, we find that $G_1$ and $G_2$ as functions of $N$ read
\begin{subequations}
\label{eq:G_1(N)_G_2(N)}
\begin{equation}
\begin{split}
    G_1 = \bigg(\cos\frac{\phi_0(N)}{2}&\cos\frac{\phi_1(N)}{2}\\
    &+ n_{01}\sin\frac{\phi_0(N)}{2}\sin\frac{\phi_1(N)}{2}\bigg)^2\,,
    \end{split}
\end{equation}
\begin{equation}
\begin{split}
    G_2 = 1& + n_{01}\sin\phi_0(N)\sin\phi_1(N)\\ +&2\bigg(\cos^2\frac{\phi_0(N)}{2}\cos^2\frac{\phi_1(N)}{2} \\
    &\qquad\qquad\quad+ n_{01}^2\sin^2\frac{\phi_0(N)}{2}\sin^2\frac{\phi_1(N)}{2}\bigg)\, ,
    \end{split}
\end{equation}
\end{subequations}
where $n_{01}\equiv\mathbf{n}_0\cdot\mathbf{n}_1$, $G_1\in[0,1]$ and $G_2\in[1,3]$. These intervals reveal that $\pi$-pulse sequences can create perfect entangling gates only within the 
{\footnotesize{CNOT}} equivalence class, where $(G_1, G_2) = (0,1)$. When the resonance condition is met, i.e., $n_{01} = -1$, we have $G_1 =\cos^2([\phi_0(N) + \phi_1(N)]/2)$. Enforcing $G_1= 0$ determines the required number of sequence iterations $N$ for achieving the wanted controlled gate. To estimate $N$, knowledge of the rotation angles in a single iteration suffices. The minima of $G_1$ occur at $N =(2k+1)\pi/(\phi_0 + \phi_1)$, however, $N$ is an integer and so the value of this expression must be rounded in general. Notice that $G_1$ can also be zero for other $N$ values, provided that $\mathrm{n}_0 \cdot \mathrm{n}_1 \leq0$.\\

\section{\label{App:Gate_fidelity_optimization}Gate fidelity optimization}
As shown in the main text, to optimize the gate fidelity up to local operations on the electronic spin, we define the target evolution operator as  
\begin{equation}
    \label{app:eq:U_0}
    \tilde{U}_0 =  R_{\mathbf{n}^e}(\theta)U_0 = \sum_{j = 0,1}(R_{\mathbf{n}^e}(\theta)\sigma_{jj})\otimes_{k=1}^KR_{\mathbf{n}_j^{(k)}}\big(\phi_j^{(k)}\big)\,,
\end{equation}
where $R_{\mathbf{n}^e}(\theta)=e^{-i\frac{\theta}{2}\bm{\sigma} \cdot\mathbf{n}^e}$ is the rotation matrix around axis $\mathbf{n}^e$ acting on the electronic spin. Next, we substitute Eqs.~\eqref{app:eq:U_0} and \eqref{eq:Kraus_operators} into Eq.~\eqref{eq:Fidelity_general_formula} to obtain
\begin{widetext}
\begin{align*}
    F &= \frac{1}{d(d+1)}\left[ {\rm tr}\left[\sum_{k=1}^{2^{L-K}}E_k^\dagger E_k\right] + \sum_{k=1}^{2^{L-K}}|{\rm tr}[\tilde{U}_0^\dagger E_k]|^2\right] = \frac{1}{d(d+1)}\left[d + \sum_k\bigg\lvert{\rm tr}\bigg[ \sum_jc_j^{(k)}p_j^{(k)}R_{\mathbf{n}^e}^\dagger(\theta)\sigma_{jj}\otimes \mathbb{1}_{2^K\times 2^K}\bigg] \bigg\rvert^2\right]\\
    &= \frac{1}{d(d+1)}\left[d + \sum_k\bigg\lvert{\rm tr}\bigg[ \sum_jc_j^{(k)}p_j^{(k)}R_{\mathbf{n}^e}^\dagger(\theta)\sigma_{jj}\bigg] {\rm tr}[\mathbb{1}_{2^K\times 2^K}] \bigg\rvert^2\right] \\
    &= \frac{1}{d(d+1)}\left[2^{K+1} + 2^{2K}\sum_k\bigg\lvert{\rm tr}\bigg[ \sum_jc_j^{(k)}p_j^{(k)}R_{\mathbf{n}^e}^\dagger(\theta)\sigma_{jj}\bigg] \bigg\rvert^2\right] = \frac{1}{2^{K+1}+1}\left[1 + 2^{K-1}\sum_k\bigg\lvert{\rm tr}\bigg[ \sum_jc_j^{(k)}p_j^{(k)}R_{\mathbf{n}^e}^\dagger(\theta)\sigma_{jj}\bigg] \bigg\rvert^2\right]\\
    &=  \frac{1}{2^{K+1}+1}\left[ 1 + 2^{K-1}\sum_k\bigg\lvert\sum_{j=0,1}c_j^{(k)}p_j^{(k)}\bigg\{\cos(\theta/2)+i(-1)^j\sin(\theta/2) n_z^{e}\bigg\}\bigg\rvert^2\right]\,, \numberthis
\end{align*}
\end{widetext}
where we used fact that $\tilde{U}_0$ is a $2^{K+1}\times 2^{K+1}$ unitary gate, the Kraus operators $E_k$ have dimensions of $2^{K+1}\times2^{K+1}$, the trace property ${\rm tr}[A\otimes B] = {\rm tr}[A]{\rm tr}[B]$, and in the end the fact that $\sigma_{jj} = \ket{j}\bra{j}$ is a projector. Also, $c_{j}^{(k)}$ and $p_{j}^{(k)}$ are given in the main text.
\section{\label{App:Success_Probability} Success Probability}
Here we present the total number of realizations we had to simulate to obtain the successful realizations used in the main text figures. As a successful realization, we define a realization in which we can find a DD sequence (CPMG or UDD) and a pair of pulse parameters $(\tau, N)$ such that a desired number of $N_{\rm R}$ nuclear spins have $\epsilon_p\geq\epsilon_p^{\rm th}$ and a desired number of $N_{\rm B}$ unwanted spins have $\epsilon_p<\epsilon_p^{\rm th}$. Any other realization is considered unsuccessful and we stop the simulation. For instance, we have an unsuccessful realization even in the case where the HF couplings are such that we cannot create a register of $N_{\rm R}$ spins surrounded by a bath of $N_{\rm B}$ spins, but we may be able to create a register of $N_{\rm R}-k$ spins surrounded by $N_{\rm B} + k$ spins (where $k$ is an integer). In Fig.~\ref{fig:Total_realizations} we show the total number of realizations needed for the silicon monovacancy, the divacancy, and the fictitious case of positive $\gamma_{^{29}\rm Si}$, where it becomes clear that the monovacancy in a low magnetic field has the highest success rate, while the fictitious bath leads to the lowest success rate. Furthermore, it becomes evident that the $(N_{\rm R}, N_{\rm B}) = (10,1)$ combination is the most challenging for all the cases, and especially for the fictitious one. However, except for the top left corner in the heatmaps, the success rate for all the other combinations is always equal to unity.
\begin{widetext}
\begin{center}
\begin{figure}[htb]
\includegraphics[width=1\textwidth,keepaspectratio]{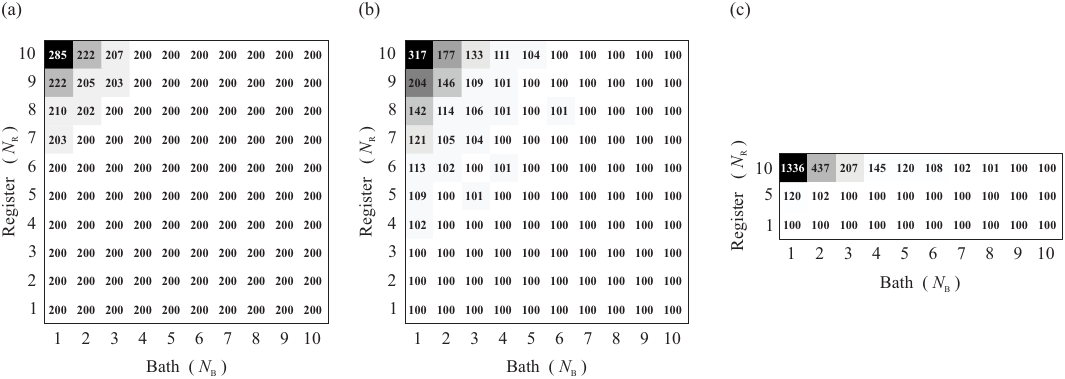}
    \caption{ \label{fig:Total_realizations}Total number of realizations simulated to obtain the results presented in the main text. Total number of realizations for (a) a silicon monovacancy defect in SiC, (b) a divacancy, and (c) a monovacancy in the fictitious nuclear spin bath where $\gamma_{^{29}\rm Si}>0$. The successful realizations for (a), (b), and (c) were 200, 100, and 100, respectively.}
\end{figure}
\end{center}
\end{widetext}
\bibliography{apssamp}% Produces the bibliography via BibTeX.
\end{document}